\documentclass[a4paper, Garamond, 12pt]{article}

\usepackage[latin1]{inputenc}
\usepackage{graphicx, amssymb, amsfonts, amsmath, mathrsfs}
\usepackage[english]{babel}
\usepackage[official]{eurosym}

\usepackage{newcent}
\usepackage{rotating}
\usepackage{multirow}
\usepackage{lineno}
\usepackage{setspace}
\usepackage{tabularx}
\usepackage[flushmargin]{footmisc} %%%%%%pas d'indentation des footnotes%%%
\usepackage{color}
\usepackage{lscape,array,hhline,pstricks,fancyhdr}
\usepackage{lineno}
\usepackage[hscale=1,vscale=1]{geometry}
\usepackage{geometry}
\usepackage{natbib}
\usepackage{url}
\usepackage{threeparttable}
\usepackage{graphicx}
\usepackage{subfigure}

\title{The asset price bubbles in emerging financial markets: a new statistical approach}
\author{Shu-Peng CHEN$^{1}$ and Ling-Yun HE$^{2, 3,}$ \footnote{Dr. HE is a full professor of applied economics and China studies. CHEN is a Ph.D. candidate supervised by Dr. HE.  This project is supported by the National Natural Science Foundation of China (Grant Nos. 71273261 and 71573258), and China National Social Science Foundation (No. 15ZDA054).}       \\ \small 1. College of Economics and Management, \\ \small China Agricultural University, Beijing 100083, China\\ \small 2. School of Economics, JiNan University, Guangzhou 510632, China\\ \small 3. School of Economics and Management, \\ \small Nanjing University of Information Science and Technology, Nanjing 210044, China\\\small * Corresponding author. \\ \small Email: lyhe@amss.ac.cn}

\date{}

\begin{document}
\maketitle
\begin{abstract}
The bubble is a controversial and important issue. Many methods which based on the rational expectation have been proposed to detect the bubble. However, for some developing countries, epically China, the asset markets are so young that for many companies, there are no dividends and fundamental value, making it difficult (if not impossible) to measure the bubbles by existing methods. Therefore, we proposed a simple but effective statistical method and three statistics (that is, C, U, V) to capture and quantify asset price bubbles, especially in immature emerging markets. To present a clear example of the application of this method to real world problems, we also applied our method to re-examine empirically the asset price bubble in some stock markets. Our main contributions to current literature are as follows: firstly, this method does not rely on fundamental value, the discount rates and dividends, therefore is applicable to the immature markets without the sufficient data of such kinds; secondly, this new method allows us to examine different influences (herding behavior, abnormal fluctuation and composite influence) of bubble. Our new statistical approach is, to the best of our knowledge, the only robust way in existing literature to to quantify the asset price bubble especially in emerging markets.
\end{abstract}



\section{Introduction}

Asset price bubble, defined as significant persistent deviations from fundamental value \citep{blanchard, blanchard1, diba1987inception, froot1991intrinsic, he2013},  is always an intensively discussed yet controversial issue, especially after the recent US subprime mortgage crisis in 2007.

What is a bubble? Is it rational or irrational? This is the simplest question and the start point of any researches on bubbles. Many theories have been proposed to answer this question, among which the widely accepted are the rational and irrational expectation bubbles. For the irrational bubbles, the scholars considered that the investors are bounded rational and tried to explain the bubbles from the phycology and behavior finance. \cite{black1986noise} firstly proposed the concept of noise traders, and based on the previous work, many models, such as noise trader models \citep{de1990noise}, popular models \citep{shiller1990speculative}, have been proposed to explain the formation of bubbles. However, it is irrelevant and pointless to insist such an ideological belief when answering such an important question. So far, both rational and irrational beliefs failed to a large extent in quantifying a bubble in real markets, especially in those immature emerging markets.

Then, another successive question arises consequently: how do we quantify a bubble? Most of these works were based on the simulation and experimental economics \citep{he2013}. However, lacking of mathematical framework lead these theories hardly to detect weather there are bubbles in real economy. In order to answer this question, the mainstream researches are focused on the rational bubble, which is based on the framework of rational expectation \citep{blanchard, blanchard1}. Many econometric methods, such as variance bounds tests \citep{shiller1981stock}, West's two-step tests \citep{west1987specification}, integration/cointegration based tests \citep{diba1987inception, diba1988theory} and intrinsic bubbles \citep{froot1991intrinsic}, etc, have been proposed to detect the existence of bubbles. However, the results  seem controversial based on these methods which are usually in ad hoc manner. Even for the same issue, different researches may give completely opposite results, just because of relaxing some assumption on the fundamentals \citep{gurkaynak2008econometric}. For example, a less restrictive fundamentals model, by allowing for time-varying discount rates, risk aversion, or structural breaks, can allow the fundamentals part of the model to fit the data better and leave less room for a bubble \citep{gurkaynak2008econometric}.

Basically, the main idea behind the existing theoretical and experimental methods is to compare the price and fundamental value. Since the fundamental value is usually unobservable, these methods were designed to study the relationship between the price and dividends. However, for some developing and emerging markets, especially China, in which the Shanghai Stock Exchange and Shenzhen Stock Exchange were set up in 1990, the capital markets are so young that the fundamental value are hardly to measure, because in these markets, many emerging companies used their profits to expand reproduction instead of giving their investors cash dividends. Meanwhile, in these emerging markets, usually the volatility is severe and huge bubble emerges in a very short period, such as 1 to 2 years, the dividend data might be unreliable and incapable in capturing the dynamical formation of a bubble. Because of these facts, the existing methods failed to a great extent to capture the character of a bubble, and to predict a bubble in real markets; furthermore, they are further challenged because of their incompetence in describing bubbles in immature emerging markets.

Based on the foresaid arguments,ignoring the conflicting (and pointless) ideological beliefs, we proposed a new robust method based on a statistical perspective, and specially tailored this method to fit the economic facts in some immature emerging markets where a fundamental value is hard or impossible to measure.

Our main contributions to current literature are as follows: Firstly, we proposed a new method, which gives us a applicable and reliable statistical perspective on bubble quantification. Secondly, this method is free of fundamental value, and does not consider the discount rates and dividends, which can be used for some emerging markets, where the fundamental value is difficult to measure, and where many listed companies have never provided any cash dividends to investors; therefore, our method could be especially applicable to quantify bubbles in some immature emerging markets without such information. And at last, we subdivided different influence (herding behavior, abnormal fluctuation and composite influence) of bubble, and test the impacts of different parts on the formation of bubble.

\section{Methods}

\subsection{Rational Bubble}

According to the theory of rational bubble, the price of a stock is constituted of two components \citep{blanchard, blanchard1}:  one is `fundamental' term $P^f$ determined by the value of the discounted dividend stream; the other is `bubble' term $B$, defined as a deviation from the fundamental or intrinsic value (see Eq.\eqref{eq:r1})

\begin{equation} \label{eq:r1}
P_t=\sum_{i=1}^{\infty} \theta^i E_t(d_{t+i})+B_t=P_t^f+B_t, \quad \text{whith} \quad \theta \equiv(1+r)^{-1}
\end{equation}
which can be solved by a utility maximization problem. Here $\theta$ stands for the discounted rate of future interest and $d$  represents the dividend.

The `bubble' part satisfies the condition

\begin{equation} \label{eq:r2}
E_t(B_{t+1})= \theta^{-1}B_t
\end{equation}
which means that investors are willing to pay higher price for the asset than the fundamental value, expecting to sell it at an even higher price in the future. If the `bubble' term increases faster (or slower) than Eq.\eqref{eq:r2},  investors can sell (or buy) this asset and obtain a higher return. The pricing of the equity is still rational; and there is no arbitrage opportunity for rational bubbles.

The return of stock price is given by
\begin{equation} \label{eq:r3}
r_t= \frac{P_{t+1}-P_t}{P_t}=\frac{(P_{t+1}^f-P_t^f)+rB_t+\varepsilon_t}{P_t}
\end{equation}
And its unconditional expectation is defined as
\begin{equation} \label{eq:r4}
E(r_t)=E(\frac{P_{t+1}^f-P_t^f}{P_t})+E(\frac{1}{P_t})[rE(B_t)+E(\varepsilon_t)]+Cov(\frac{1}{P_t},rB_t+\varepsilon_t)
\end{equation}

If  $B_t$ is non-zero, the rational bubble exists in the asset price. It is clear from Eq.\eqref{eq:r2} that this bubble is non-stationary, and increases at the geometric rate $\theta^{-1}$. If there is no bubble, the solution is $B_t=0$, namely, the value of bubble term is always zero. Then we can get
\begin{equation} \label{eq:r5}
E(r_t)=E(\frac{P_{t+1}^f-P_t^f}{P_t})+E(\frac{1}{P_t})E(\varepsilon_t)+Cov(\frac{1}{P_t},\varepsilon_t)
\end{equation}
The  ${\varepsilon _t}$ is a random variable generated by a stochastic process, namely, it is independent of the price $P_t$. The expectation of bubble increases at the ratio $\theta^{-1}$, namely,${E_t}\left( {{B_{t + 1}}} \right) = {\theta ^{ - 1}}E\left( {{B_t}} \right)$. It is obvious that $E\left( {{\varepsilon _t}} \right) = 0$. Then Eq.\eqref{eq:r5} can be rewritten by
\begin{equation} \label{eq:r6}
E(r_t)=E(\frac{P_{t+1}^f-P_t^f}{P_t})
\end{equation}

Eq.\eqref{eq:r6} implies that if there is no bubble in asset price, the return is relative to the change of fundamental value, namely, is determined by $P_{t + 1}^f - P_t^f$. For example, if an investor buys the stock of a company, she will not only obtain the dividend, but also get the return of capital appreciation, which can fully reflect the company's growth in the future.

The fundamental value $P^f$ is determined by the discounted dividend stream in future; however, the dividend is usually paid once only a year. If the time interval $[t, t+1]$ is short enough, say, one minute, it is reasonable to assume that the fundamental value of the stock would remain the same within a very short time interval, namely, $P_t^f = P_{t + 1}^f$. Then Eq.\eqref{eq:r6} can be rewritten as
\begin{equation} \label{eq:r7}
E(r_t)=0
\end{equation}

Eq.\eqref{eq:r7} implies that if there is no bubble and if the fundamental value does not change, the behavior of stock would be a fair game that nobody can get the additional profits; furthermore, the daily returns should be independent under these circumstances. Suppose that if the returns are not independent, a  rational investor can recognize and take advantage of the chance, and thereby get additional profits. The rational investor's behaviors will affect the price; therefore, Eq.\eqref{eq:r7} is not satisfied. Note that within a short-enough time interval $[t, t+1]$, the fundamental value will remain the same; then the deviation only come from the bubble term $B_t$ in the pricing equation (Eq.\eqref{eq:r1}), namely, $B_t \neq 0$, which contradicts the condition of no bubble. Then if there is no bubble, the daily returns should be independent; in other words, the price should be random walk and the market should be at least weak-form efficient according to Fama's Efficient Market Hypothesis \citep{Fama}.

If the returns are $i.i.d$ with finite variances, according to the Central Limit Theorem \citep[pp. 112-113]{Spiegel}, they are normal distributed. However, numerous empirical studies showed that in financial markets, the distribution is not normal, showing a high degree of peakedness and fat tails \citep{Mandelbrot1}. Meanwhile, the recent researches also show some ubiquitous properties (e.g. long-term correlation, fractality) which reject the random walk hypothesis  \citep{Mandelbrot2}. As we know, if there is no bubble, the returns should be independent; otherwise, the returns should not be independent. For a simplest example, if there is a bubble, an investor can predict the trend of price by the history information, because there is a high probability for the price to rise even higher. Eq.\eqref{eq:r7} should be rewritten by $E\left( {{r_t}} \right) = rE\left( {{{{B_t}} \mathord{\left/ {\vphantom {{{B_t}} {{P_t}}}} \right. \kern-\nulldelimiterspace} {{P_t}}}} \right) > 0$. The returns in a bubble are higher than those without bubble. If one use the whole length of time series, in some periods of which there may exist bubbles, while in other periods of which there is no bubble, the bubble will push the distribution away from normality, and exhibit fat tails. Therefore, one can detect the bubble according to the degree of deviation from random walk and normal distribution.

\subsection{Bubble quantification}

If the price contains all  information available and the market is efficient,  then bubble cannot exist and asset price  behavior follows random walk, namely, the price will rise or fall with the equal probability in the next time interval. However, in the real market, the cases are totally different. Herd behaviors among the investors would push the price further deviating from the fundamental value. The price will rise even higher, and the upside probability is larger than that of downside. In order to capture this feature in real markets, we introduce the function
\begin{equation} \label{eq:r8}
u(r_t) = \begin{cases}
              1, \quad \text{if} \quad r_t>0 \\
              0, \quad \text{if} \quad r_t \leq 0
             \end{cases}
\end{equation}

And for the time series of daily returns, we incorporate a window box with size of $N$  and calculate the moving average function
\begin{equation} \label{eq:r9}
U_t=\frac{1}{N}\sum_{i=-[N/2]}^{N-[N/2]-1} u(r_{t+i})
\end{equation}
where $[x]$ represents the largest integer less than $x$. Furthermore, consider that $r: N\left( {0,{\sigma ^2}} \right)$, then the unconditional expectation $ E\left( U \right) = \frac{1}{2}$  and variance $Var\left( U \right) = \frac{1}{{4N}}$ (See Appendix A). If there is no bubble which is caused by the herd behavior, namely, the upside and downside probabilities are the same, the statistic $U$  should be subject to the binomial distribution, or according to the central limit theorem, it is also can be considered as normal distribution. We considered to use the hypothesis test to detect the existence of bubble, namely, the null hypothesis $H_0: E\left( U \right) = \frac{1}{2}$.

Meanwhile, given the existence of price bubble, there exist not only the asymmetric upside and downside probability, but also the asymmetric price change. It can be considered from the two aspects: First, from Eq.\eqref{eq:r2} we can find that it is a power-low growing bubble, the magnitude of price rising is larger than that of falling when the bubble grows; Second, for the burst of a bubble, especially for the crash, the panic of investors will lead the price drop quickly, and the downside risk is much larger than that of upside. Based on this influence, we designed the statistic $V$  to examine the asymmetry of price change. Considering a window box with size of $N$, we noted all of the positive returns in the box as a set $S_t^+ $ ($S_t^+  = \left\{ {{r_i}|{r_i} > 0,t - \left[ {{N \mathord{\left/ {\vphantom {N 2}} \right. \kern-\nulldelimiterspace} 2}} \right] \le i \le t + N - \left[ {{N \mathord{\left/ {\vphantom {N 2}} \right. \kern-\nulldelimiterspace} 2}} \right] - 1} \right\}$), in which the element ${r^+ } \in S_t^+ $ and the number of element was noted as
$N_t^ + $; meanwhile, the other non-positive returns were constituted as a set $S_t^ - $ ($S_t^ -  = \left\{ {{r_i}|{r_i} \le 0,t - \left[ {{N \mathord{\left/ {\vphantom {N 2}} \right.\kern-\nulldelimiterspace} 2}} \right] \le i \le t + N - \left[ {{N \mathord{\left/ {\vphantom {N 2}} \right. \kern-\nulldelimiterspace} 2}} \right] - 1} \right\}$), in which the element ${r^ - } \in S_t^ - $  and the number of element was noted as $N_t^ - $. Then we designed the statistic $V$:
\begin{equation} \label{eq:r10}
V_t=\frac{1}{N_t^+}\sum_{r^+ \in S_t^+} r^+ + \frac{1}{N_t^-}\sum_{r^- \in S_t^-} r^-
\end{equation}

If the time series is the random walk, namely there is no bubble, the unconditional expectation $E\left( {{V_t}} \right) = 0$ and variance $Var\left( {{V_t}} \right) = \left( {\frac{1}{{{N^ + }}} + \frac{1}{{{N^ - }}}} \right)\frac{{\pi  - 2}}{\pi }{\sigma ^2}$  (See Appendix A) and according to the central limit theorem, the statistic $V$  should be subject to the normal distribution. In order to examine the existence of bubble, we also used the null hypothesis  $H_0: E\left( V \right) = 0$ .

At last, in order to avoid the situation if both of the two methods failed, we also designed the composite statistic $C$
\begin{equation} \label{eq:r11}
C_t=\frac{1}{N_t^+}\sum_{r^+ \in S_t^+} r^+\frac{1}{N}\sum_{i=-[N/2]}^{N-[N/2]-1} u(r_{t+i}) +\frac{1}{N_t^-}\sum_{r^- \in S_t^-} r^-\frac{1}{N}\sum_{i=-[N/2]}^{N-[N/2]-1}(1-u(r_{t+i}))
\end{equation}

We can find that unconditional expectation $E(C_t)=0$ and variance $Var(C_t)=[\frac{(\pi-2)(N+1)}{4\pi N^+ N^-}+\frac{2}{N\pi}]\sigma^2$ (See Appendix A); meanwhile, the statistic $C$ is approximately subject to normal distribution (See Appendix B). Then in order to detect the bubble, we use the null hypothesis $H_0 : E(C)=0$.

\begin{figure}[h!]
\begin{center}
\includegraphics[height=3in]{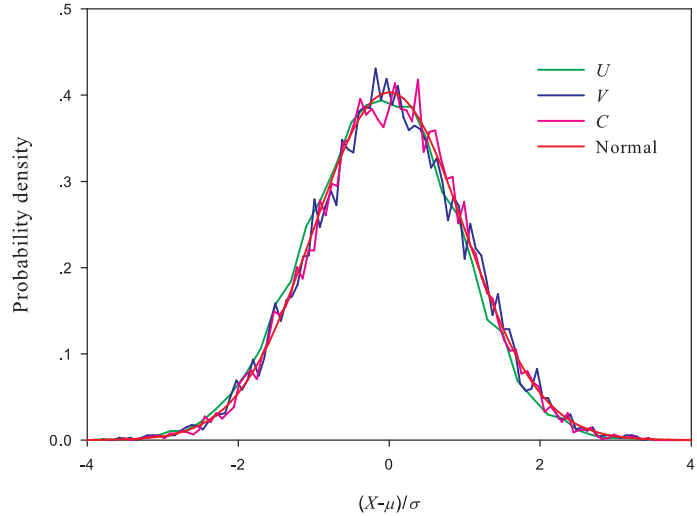}
\caption{The probability density of the statistics $U$ (green line), $V$ (blue line) and $C$ (pink line) from random walk time series. The red line is the probability density of standard normal distribution.} \label{fig1}
\end{center}
\end{figure}

All of our three designed statistics are based on the normal distribution. in order to test the validity, we generate the random number $r$ with the length of $N$\footnote{Here we chose $N=100$}, which $r\sim N(0,1)$, and then calculate the statistics $U$, $V$ and $C$. We repeat it $10000$ times and get the distribution. Fig. \ref{fig1} shows the probability density of normalized statistics $U$ (green line), $V$ (blue line) and $C$ (pink line). As a contrast, we also give the standard normal distribution (red line), from which we could clearly find that all of the three statistics were subject to the normal distribution.

\section{Empirical Study}

\subsection{Data}

In order to study the bubble both in Chinese and US stock markets, we used the daily closing prices of Shanghai Composite Index (SHCI for short) from January 4th, 2005 to December 12th, 2012 (the length of data is 1931) and Dow Jones Industrial Average (DJIA for short) from January 4th, 2005 to December 12th, 2012 (the length of data is 1999).

To get a better understanding of the data sets, we provided the summary statistics of the daily returns from the two indexes. The sknewness and kurtosis show the skewed and leptokurtic distribution in both of the two markets, meanwhile, the Jarque-Bera test shows that it reject the null hypothesis of normal distribution at the significance of $0.01$. All of these suggested that they were not subject to the normal distribution, however, just as we have analyzed previously, even if the market was effective and followed the behavior of random walk, the emergence of bubble would drive it away from normal.

\begin{table}[h!]
\centering
\begin{threeparttable}
\caption{The summary statistics of the daily returns from SHCI and DJIA} \label{tab1}
\begin{tabular}{c*{6}{c}}
\hline
     &   Mean & {Std. dev.} & Skewness & Kurtosis & Jarque-Bera\\\hline
SHCI & 0.0004 &      0.0178 &  -0.1771 &   6.1827 & 824.65*\\
DJIA & 0.0002 &      0.0128 &   0.1724 &  13.3386 & 8908.3*\\
\hline
\end{tabular}
\begin{tablenotes}
   \footnotesize
   \item[*] Means reject the null hypothesis that the sample comes from a normal distribution at the significance of 0.01.
\end{tablenotes}
\end{threeparttable}
\end{table}

\subsection{Bubble Detection}

Before our calculation, we should select the window box $N$ carefully. The $N$ could not be too small, because small $N$ might lead the loss of statistical properties; meanwhile, the $N$ could not be too large, because in our assumption, during this time, the fundamental value of company would not change. Based on the above considerations, we chose $N=100$.

Another problem is the standard error $\sigma$ of the returns $r$. The variance of $V$ and $C$ are based on it, however, we don't know what it is. If the time series is just a random walk, $S^2=\frac{1}{L-1}\sum(r-\bar{r})$ is the unbiased estimation of $\sigma^2$. But if there is a bubble in the price, especially the power-growth bubble, the standard error which we estimated from the sample would be larger than the real $\sigma$. Considering that if the returns are independent and subject to the normal distribution, $99.7\%$ data of sample would fall in the range of $3$ standard error. Then we estimated the standard error according to following steps:

1) Calculating the standard error from the sample;

2) Excluding the data which are larger than $3$ standard error;

3) Calculating the standard error from the remaining sample.

Through the process, we can reduce the influence of the bubble (if it exists). However, if there is no bubble, this estimation would be a little smaller than the real $\sigma$. Fortunately, we found that this process is approximately equivalent to estimate from a truncated distribution, and the estimation $E(S^2\mid|r|\leq 3\sigma)=K\sigma^2$, where $K=1-6\exp(-9/2)/(\sqrt{2\pi}(2\Phi(3)-1))$ and the $\Phi(*)$ means the cumulative distribution function of standard normal distribution (see Appendix C). By diving the $\sqrt{K}$, the estimation is unbiased.

\begin{figure}[h!]
\begin{center}
\includegraphics[height=3in,width=4in]{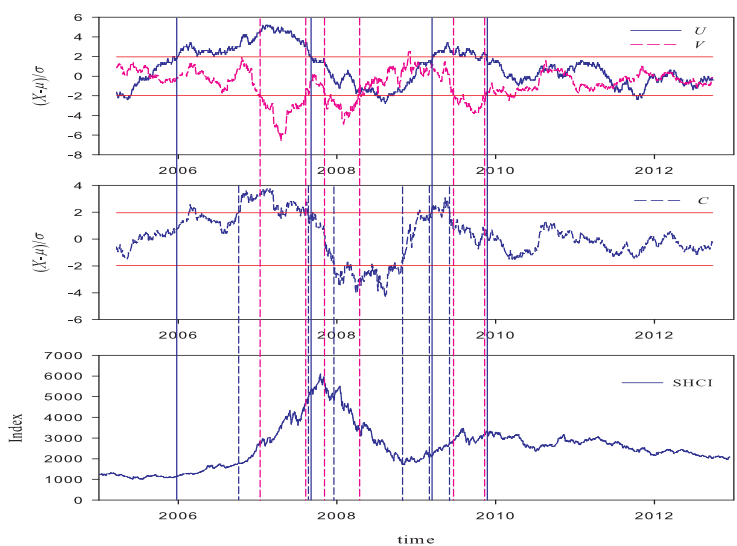}
\caption{The normalized statistics $U$, $V$ (upper plot) and $C$ (middle plot) calculated from the Shanghai Composite Index. The lower plot gives the index. The red lines mean the $95\%$ confidence interval. The plot also shows the periods of $U$ (vertical blue solid line), $V$ (vertical pink dotted line) and $C$ (vertical blue dotted line) which exceed the $95\%$ confidence interval.} \label{fig2}
\end{center}
\end{figure}

Fig. \ref{fig2} shows the results of SHCI under the confidence level $\alpha=0.05$. The plots shows the normalized statistics $U$, $V$ (upper plot) and $C$ (middle plot). The red lines mean the $95\%$ confidence interval. Meanwhile, we also give the periods of $U$ (vertical blue solid line), $V$ (vertical pink dotted line) and $C$ (vertical blue dotted line) which exceed the $95\%$ confidence interval. The statistic $U$ is used to measure the rising probability of the price. If the bubble booms, it is easy for the price to rise even higher than before. The investors bought the stock because they thought that they could sell it with higher price in the future, which was based on the rational expectation. Their similar behaviors would promote the bubble continues to bloom, or in other words, they were also called the imitation or herding behaviors. Based on this case, it is not a fair game. The price is non-stationary and there is a higher rising probability. The statistic $U$ is designed to measure this phenomenon and the results of SHCI have been shown in Table \ref{tab2}. From the table we can find that there are two periods which exceed the $95\%$ confidence interval, namely, from Dec.7th, 2005 to Sep. 3rd, 2007, and from Mar. 17th, 2009 to Nov. 18th, 2009\footnote{This period does not mean the start and end time of the bubble. It only shows that in this period the bubble is detected.}, which is corresponding to the huge bubble during 2005 to 2007 and a small bubble blooming in 2009. Meanwhile, we also show the normalized max value of $U$ during these periods and their P value (see Table \ref{tab2}). The statistic $V$ is used to measure the abnormal fluctuations, especial for the crashes. If the bubble bursts, it is easier for the investors to become panic. The sellers eager to sell their assets, however, in this moment there are only a few buyers. In the short period, the sellers cannot find the buyers in current price and they can only quote an even lower price. The statistic $V$ is designed to measure this phenomenon and it can be also treated as a psychological factor. From the SHCI, we can find that there are three periods which exceed the $95\%$ confidence interval, two of which are from Nov. 7th, 2007 to May. 5th, 2008, and from Jun. 19th, 2009 to Nov. 13th, 2009 that is corresponding to the bubble bursting in 2007 and 2009. More interesting, we also find that there is a period of abnormal fluctuation from Jan. 9th, 2007 to Aug. 22nd, 2007, which is during the time of bubble blooming in 2007. This means that during 2007, there was a short period of bursting, however, there was not a crash, instead, the bubble continued to bloom. The statistic $C$ is designed as a composite index, which combines the influence of $U$ and $V$, and applied as a supplementary method to measure the bubble. From the table, we can find that there are three periods, namely, from Sep. 29th, 2006 to Aug. 17th, 2007, from Dec. 19th, 2007 to Oct. 23rd, 2008, and from Jan. 8th, 2009 to Jun. 1st, 2009, which confirmed the Chinese stock bubble in 2007 and 2009.

\begin{figure}[h!]
\begin{center}
\includegraphics[height=3in,width=4in]{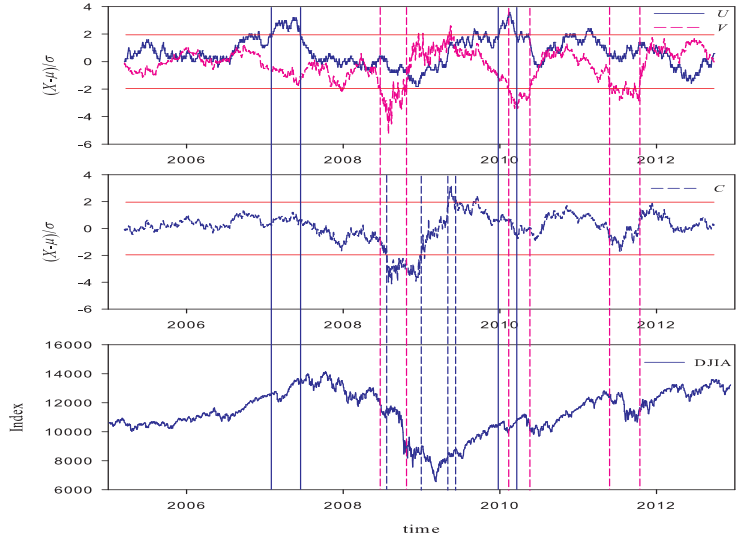}
\caption{The normalized statistics $U$, $V$ (upper plot) and $C$ (middle plot) calculated from the Dow Jones Industrial Average. The lower plot gives the index. The red lines mean the $95\%$ confidence interval. The plot also shows the periods of $U$ (vertical blue solid line), $V$ (vertical pink dotted line) and $C$ (vertical blue dotted line) which exceed the $95\%$ confidence interval.} \label{fig3}
\end{center}
\end{figure}

Fig. \ref{fig3} shows the results of DJIA under the confidence level $\alpha=0.05$ and Table \ref{tab3} shows the periods which exceed the $95\%$ confidence interval. Meanwhile, we also give the normalized max(min) value during periods and their P value. From the table, we can find that, for the statistic $U$, there are two periods, one is from Jan. 24th, 2007 to Jun. 14th, 2007, which is during the US subprime crisis, another is from Mar. 17th, 2009 to Nov. 18th, 2009, which means that since 2009, the bubble became to bloom; from the statistic $V$, there are three periods, one among them is from Jun. 23rd, 2008 to Oct. 23rd, 2008, which is corresponding to the crashes in 2008, and another two periods are from Feb. 11th, 2010 to May. 19th, 2010, and from Jun. 7th, 2011-Oct. 12th, 2011, which mean that there are two times of bubble bursting during this period. However, it is confused that after the bursting, the price rose even higher, which is similar with the Chinese stock bubble in 2007 (see Fig. \ref{fig2}). This may mean that the bubble still exists and continues to bloom, however, the speed is not so fast that the statistic is approaching but not exceeding the $95\%$ confidence interval; from the statistic $C$, there are two periods, one is from Jul. 17th, 2008 to Dec. 30th, 2008, which is corresponding the bubble bursting during 2008, and another period is from May. 7th, 2009 to Sep. 22nd, 2009, which confirmed that the bubble became to bloom since 2009.

\begin{table}[h!]
\centering
\caption{the results of SHCI} \label{tab2}
\begin{tabular}{c*{4}{c}}
\hline
      Statistics &                         Periods & Max(Min) Value & P Value\\\hline
\multirow{2}*{U} &    Dec.7th, 2005-Sep. 3rd, 2007 &         5.2000 & 0.0000\\
                 & Mar. 17th, 2009-Nov. 18th, 2009 &         3.4000 & 0.0003\\\cline{2-4}
\multirow{3}*{V} &  Jan. 9th, 2007-Aug. 22nd, 2007 &        -6.6329 & 0.0000\\
                 &   Nov. 7th, 2007-May. 5th, 2008 &        -4.8869 & 0.0000\\
                 & Jun. 19th, 2009-Nov. 13th, 2009 &        -3.8687 & 0.0001\\\cline{2-4}
\multirow{3}*{C} & Sep. 29th, 2006-Aug. 17th, 2007 &         3.7625 & 0.0001\\
                 & Dec. 19th, 2007-Oct. 23rd, 2008 &        -4.2667 & 0.0000\\
                 &   Jan. 8th, 2009-Jun. 1st, 2009 &         3.0811 & 0.0010\\
\hline
\end{tabular}
\end{table}

\begin{table}[h!]
\centering
\caption{the results of DJIA} \label{tab3}
\begin{tabular}{c*{4}{c}}
\hline
      Statistics &                         Periods & Max(Min) Value & P Value\\\hline
\multirow{2}*{U} & Jan. 24th, 2007-Jun. 14th, 2007 &         3.2000 & 0.0007\\
                 & Dec. 28th, 2009-Mar. 17th, 2010 &         3.6000 & 0.0002\\\cline{2-4}
\multirow{3}*{V} & Jun. 23rd, 2008-Oct. 23rd, 2008 &        -5.2396 & 0.0000\\
                 & Feb. 11th, 2010-May. 19th, 2010 &        -3.3992 & 0.0003\\
                 &  Jun. 7th, 2011-Oct. 12th, 2011 &        -2.8524 & 0.0022\\\cline{2-4}
\multirow{2}*{C} & Jul. 17th, 2008-Dec. 30th, 2008 &        -4.1298 & 0.0000\\
                 &  May. 7th, 2009-Sep. 22nd, 2009 &         3.1903 & 0.0007\\
\hline
\end{tabular}
\end{table}

\section{Conclusions}

We proposed a new method to detect the bubble, which gives us a new perspective to detect the bubble. does not rely on fundamental value, the discount rates and dividends, therefore is applicable to the immature markets without the sufficient data of such kinds; secondly, this new method allows us to examine different influences (herding behavior, abnormal fluctuation and composite influence) of bubble. Comparing with the existing methods, our method  also has the additional advantage of applying daily data instead of monthly or quarterly data, thereby including more information,

To present a clear example of the application of this method to real world problems, we also applied our method to re-examine empirically the asset price bubble in some stock markets. Our results show that for Chinese stock market, the results confirmed the bubble in 2007 and 2009; meanwhile, for US stock market, it successfully detected the bubble during the subprime crisis.

Our new statistical approach is, to the best of our knowledge, the only and robust way in existing literature to to quantify the asset price bubble especially in emerging markets. Our endeavor would bring more insights and better understandings in asset price bubble, and provide a new approach to quantify bubble in real world markets from a simple but effective statistical perspective.

%\section*{References}
\bibliographystyle{elsarticle-harv}
\bibliography{MyReference}

\begin{thebibliography}{15}
\expandafter\ifx\csname natexlab\endcsname\relax\def\natexlab#1{#1}\fi
\providecommand{\url}[1]{\texttt{#1}}
\providecommand{\href}[2]{#2}
\providecommand{\path}[1]{#1}
\providecommand{\DOIprefix}{doi:}
\providecommand{\ArXivprefix}{arXiv:}
\providecommand{\URLprefix}{URL: }
\providecommand{\Pubmedprefix}{pmid:}
\providecommand{\doi}[1]{\href{http://dx.doi.org/#1}{\path{#1}}}
\providecommand{\Pubmed}[1]{\href{pmid:#1}{\path{#1}}}
\providecommand{\bibinfo}[2]{#2}
\ifx\xfnm\relax \def\xfnm[#1]{\unskip,\space#1}\fi
%Type = Article
\bibitem[{Black(1986)}]{black1986noise}
\bibinfo{author}{Black, F.}, \bibinfo{year}{1986}.
\newblock \bibinfo{title}{Noise}.
\newblock \bibinfo{journal}{Journal of Finance} \bibinfo{volume}{41},
  \bibinfo{pages}{529--543}.
%Type = Article
\bibitem[{Blanchard and Watson(1983)}]{blanchard}
\bibinfo{author}{Blanchard, O.}, \bibinfo{author}{Watson, M.},
  \bibinfo{year}{1983}.
\newblock \bibinfo{title}{Bubbles, rational expectations and financial
  markets}.
\newblock \bibinfo{journal}{NBER Working Paper} .
%Type = Article
\bibitem[{Blanchard(1979)}]{blanchard1}
\bibinfo{author}{Blanchard, O.J.}, \bibinfo{year}{1979}.
\newblock \bibinfo{title}{Speculative bubbles, crashes and rational
  expectations}.
\newblock \bibinfo{journal}{Economics letters} \bibinfo{volume}{3},
  \bibinfo{pages}{387--389}.
%Type = Article
\bibitem[{Chen and He(2013)}]{he2013}
\bibinfo{author}{Chen, S.P.}, \bibinfo{author}{He, L.Y.}, \bibinfo{year}{2013}.
\newblock \bibinfo{title}{Bubble formation and heterogeneity of traders: A
  multi-agent perspective}.
\newblock \bibinfo{journal}{Computational Economics} \bibinfo{volume}{42},
  \bibinfo{pages}{267--289}.
%Type = Article
\bibitem[{De~Long et~al.(1990)De~Long, Shleifer, Summers and
  Waldmann}]{de1990noise}
\bibinfo{author}{De~Long, J.B.}, \bibinfo{author}{Shleifer, A.},
  \bibinfo{author}{Summers, L.H.}, \bibinfo{author}{Waldmann, R.J.},
  \bibinfo{year}{1990}.
\newblock \bibinfo{title}{Noise trader risk in financial markets}.
\newblock \bibinfo{journal}{Journal of Political Economy} \bibinfo{volume}{98},
  \bibinfo{pages}{703--738}.
%Type = Article
\bibitem[{Diba and Grossman(1987)}]{diba1987inception}
\bibinfo{author}{Diba, B.T.}, \bibinfo{author}{Grossman, H.I.},
  \bibinfo{year}{1987}.
\newblock \bibinfo{title}{On the inception of rational bubbles}.
\newblock \bibinfo{journal}{Quarterly Journal of Economics}
  \bibinfo{volume}{102}, \bibinfo{pages}{697--700}.
%Type = Article
\bibitem[{Diba and Grossman(1988)}]{diba1988theory}
\bibinfo{author}{Diba, B.T.}, \bibinfo{author}{Grossman, H.I.},
  \bibinfo{year}{1988}.
\newblock \bibinfo{title}{The theory of rational bubbles in stock prices}.
\newblock \bibinfo{journal}{The Economic Journal} \bibinfo{volume}{98},
  \bibinfo{pages}{746--754}.
%Type = Article
\bibitem[{Froot and Obstfeld(1991)}]{froot1991intrinsic}
\bibinfo{author}{Froot, K.A.}, \bibinfo{author}{Obstfeld, M.},
  \bibinfo{year}{1991}.
\newblock \bibinfo{title}{Intrinsic bubbles: The case of stock prices}.
\newblock \bibinfo{journal}{American Economic Review} \bibinfo{volume}{81 (5)},
  \bibinfo{pages}{1189--1214}.
%Type = Article
\bibitem[{G{\"u}rkaynak(2008)}]{gurkaynak2008econometric}
\bibinfo{author}{G{\"u}rkaynak, R.S.}, \bibinfo{year}{2008}.
\newblock \bibinfo{title}{Econometric tests of asset price bubbles: Taking
  stock}.
\newblock \bibinfo{journal}{Journal of Economic Surveys} \bibinfo{volume}{22},
  \bibinfo{pages}{166--186}.
%Type = Article
\bibitem[{Mandelbrot(1963)}]{Mandelbrot1}
\bibinfo{author}{Mandelbrot, B.B.}, \bibinfo{year}{1963}.
\newblock \bibinfo{title}{The variation of certain speculative prices}.
\newblock \bibinfo{journal}{Journal of Business} \bibinfo{volume}{36 (4)},
  \bibinfo{pages}{394--419}.
%Type = Article
\bibitem[{Mandelbrot(1971)}]{Mandelbrot2}
\bibinfo{author}{Mandelbrot, B.B.}, \bibinfo{year}{1971}.
\newblock \bibinfo{title}{When can price be arbitraged efficiently? a limit to
  the validity of the random walk and martingale models}.
\newblock \bibinfo{journal}{Review of Economics and Statistics}
  \bibinfo{volume}{53 (3)}, \bibinfo{pages}{225--236}.
%Type = Article
\bibitem[{Shiller(1981)}]{shiller1981stock}
\bibinfo{author}{Shiller, R.J.}, \bibinfo{year}{1981}.
\newblock \bibinfo{title}{Do stock prices move too much to be justified by
  subsequent changes in dividends?}
\newblock \bibinfo{journal}{The American Economic Review} \bibinfo{volume}{71},
  \bibinfo{pages}{421--436}.
%Type = Article
\bibitem[{Shiller(1990)}]{shiller1990speculative}
\bibinfo{author}{Shiller, R.J.}, \bibinfo{year}{1990}.
\newblock \bibinfo{title}{Speculative prices and popular models}.
\newblock \bibinfo{journal}{Journal of Economic Perspectives}
  \bibinfo{volume}{4}, \bibinfo{pages}{55--65}.
%Type = Book
\bibitem[{Spiegel(1992)}]{Spiegel}
\bibinfo{author}{Spiegel, M.R.}, \bibinfo{year}{1992}.
\newblock \bibinfo{title}{Theory and Problems of Probability and Statistics}.
\newblock \bibinfo{publisher}{New York: McGraw-Hill}.
%Type = Article
\bibitem[{West(1987)}]{west1987specification}
\bibinfo{author}{West, K.D.}, \bibinfo{year}{1987}.
\newblock \bibinfo{title}{A specification test for speculative bubbles}.
\newblock \bibinfo{journal}{Quarterly Journal of Economics}
  \bibinfo{volume}{102}, \bibinfo{pages}{553--580}.

\end{thebibliography}

\clearpage

\appendix

\section{The Mean and Variance of the Statistics $U$, $V$ and $C$}

The Statistic $U$ is designed to examine the probability of price rising. If $r$ is positive, the function $u(r)=1$, while if $r$ is non-positive, $u(r)=0$. If there is no bubble in the price and the market is effective, there is equal probability for the asset price to rise or fall in the tomorrow. It is obvious that the expectation $E(u(r))=\frac{1}{2}$ and the variance $Var(u(r))=\frac{1}{4}$.

Then the expectation of $U$

\begin{equation*}
E(U)=E\left(\frac{1}{N}\sum_{i=1}^N u(r_i)\right)=E(u(r))=\frac{1}{2}
\end{equation*}
And the variance

\begin{equation*}
Var(U)=Var\left(\frac{1}{N}\sum_{i=1}^N u(r_i)\right)=\frac{1}{N}Var(u(r))=\frac{1}{4N}
\end{equation*}

The Statistic $V$ is used to examine the magnitude of price change. For an effective market, the return $r$ is independent and subject to normal distribution with mean $0$ and variance $\sigma^2$, namely, $r\sim N(0,\sigma^2)$. For a period of time series, the $r^+$ presidents the positive return $r$, then the cumulative distribution function (cdf)

\begin{align*}
F(r^+) & =F(r\mid r>0)\\
       & =P\{R\leq r\mid r>0\}\\
       & =\frac{P\{0<R\leq r\}}{P\{r>0\}}\\
       & =\begin{cases}
             2\int_0^r f(r) dr, & \text{if} \quad r>0\\
             0, & \text{if} \quad r\leq 0\\
          \end{cases}
\end{align*}

And the probability density function (pdf)

\begin{equation*}
f(r^+)=\begin{cases}
          2f(r), & \text{if} \quad r>0\\
          0, & \text{if} \quad r\leq 0\\
       \end{cases}
\end{equation*}

Then we can obtain the expectation of  $r^+$

\begin{align*}
E(r^+) & =\int_{-\infty}^{+\infty} r^+ f(r^+)dr^+\\
       & =2\int_0^{+\infty} rf(r)dr\\
       & =2\int_0^{+\infty} r\frac{1}{\sqrt{2\pi}\sigma}e^{-\frac{r^2}{2\sigma^2}}dr\\
       & =\sqrt{\frac{2}{\pi}}\sigma\\
\end{align*}

And the variance

\begin{align*}
Var(r^+) & =E\left[(r^+)^2\right]-\left[E(r^+)\right]^2\\
         & =2\int_0^{+\infty} r^2f(r)dr-\frac{2}{\pi}\sigma^2\\
         & =2\int_0^{+\infty} r^2\frac{1}{\sqrt{2\pi}\sigma}e^{-\frac{r^2}{2\sigma^2}}dr-\frac{2}{\pi}\sigma^2\\
         & =\frac{\pi-2}{\pi}\sigma^2
\end{align*}
Similarly, we can obtain the expectation $E(r^-)=-E(r^+)=-\sqrt{\frac{2}{\pi}}\sigma$ and the variance $Var(r^-)=Var(r^+)=\frac{\pi-2}{\pi}\sigma^2$.

Then, for the Statistic $V$, the expectation

\begin{align*}
E(V) & =E\left(\frac{1}{N^+}\sum_{r^+ \in S^+} r^+ + \frac{1}{N^-}\sum_{r^- \in S^-} r^- \right)\\
     & =E(r^+)+E(r^-)\\
     & =0\\
\end{align*}
It is obvious that $r^+$ and $r^-$ are independent, and then we can obtain the variance

\begin{align*}
Var(V) & =Var\left(\frac{1}{N^+}\sum_{r^+ \in S^+} r^+ + \frac{1}{N^-}\sum_{r^- \in S^-} r^- \right)\\
       & =\frac{1}{N^+}Var(r^+)+\frac{1}{N^-}Var(r^-)\\
       & =\left(\frac{1}{N^+}+\frac{1}{N^-}\right)\frac{\pi-2}{\pi}\sigma^2
\end{align*}

For each random variable, the absolute value and the symbol are independent, namely, the $r^+$, $r^-$ and $u(r)$ are independent. Then, for the statistic $C$, we can get the expectation

\begin{align*}
E(C) & =E\left[\frac{1}{N^+}\sum_{r^+ \in S^+} r^+\frac{1}{N}\sum_{i=1}^N u(r_i) +\frac{1}{N^-}\sum_{r^- \in S^-} r^-\frac{1}{N}\sum_{i=1}^N(1-u(r_i))\right]\\
     & =E(r^+)E(u(r))+E(r^-)\left[1-E(u(r))\right]\\
     & =0\\
\end{align*}

And the variance

\begin{align*}
Var(C) & =E(C^2)-\left[E(C)\right]^2\\
       & =E\left[\left(\frac{1}{N^+}\sum_{r^+ \in S^+} r^+\frac{1}{N}\sum_{i=1}^N u(r_i) +\frac{1}{N^-}\sum_{r^- \in S^-} r^-\frac{1}{N}\sum_{i=1}^N(1-u(r_i))\right)^2\right]\\
       & =\left(\frac{1}{N^+}\right)^2E\left[\left(\sum_{r^+ \in S^+} r^+\right)^2\right] E\left[\left(\frac{1}{N}\sum_{i=1}^N u(r_i)\right)^2\right]+ \left(\frac{1}{N^-}\right)^2E\left[\left(\sum_{r^- \in S^-} r^-\right)^2\right] E\left[\left(\frac{1}{N}\sum_{i=1}^N (1-u(r_i))\right)^2\right]\\
       & +\frac{2}{N^+ N^-}E\left(\sum_{r^+ \in S^+} r^+\right)E\left(\sum_{r^- \in S^-} r^-\right)E\left[\frac{1}{N^2}\sum_{i=1}^N u(r_i)\sum_{i=1}^N (1-u(r_i))\right]\\
\end{align*}
For each part of the equation, we can calculate respectively,

\begin{align*}
E\left[\left(\sum_{r^+ \in S^+} r^+\right)^2\right] & =Var\left(\sum_{r^+ \in S^+} r^+\right)+\left[E\left(\sum_{r^+ \in S^+} r^+\right)\right]^2\\
                                                    & =\sum_{r^+ \in S^+}Var(r^+)+\left[\sum_{r^+ \in S^+}E(r^+)\right]^2\\
                                                    & =N^+ \frac{\pi-2+2N^+}{\pi}\sigma^2\\
\end{align*}
Similarly, $E\left[\left(\sum_{r^- \in S^-} r^-\right)^2\right]=N^- \frac{\pi-2+2N^-}{\pi}\sigma^2$,

\begin{align*}
E\left[\left(\frac{1}{N}\sum_{i=1}^N u(r_i)\right)^2\right] & =\frac{1}{N^2}E\left[\left(\sum_{i=1}^N u(r_i)\right)^2\right]\\
                                                            & =\frac{1}{N^2}\left[Var\left(\sum_{i=1}^N u(r_i)\right)+\left(E\left(\sum_{i=1}^N u(r_i)\right)\right)^2\right]\\
                                                            & =\frac{1+N}{4N}\\
\end{align*}
Similarly,

\begin{align*}
E\left[\left(\frac{1}{N}\sum_{i=1}^N (1-u(r_i))\right)^2\right] & =\frac{1}{N^2}E\left[N^2-2N\sum_{i=1}^N u(r_i)+\left(\sum_{i=1}^N u(r_i)\right)^2\right]\\
                                                                & =1-\frac{2}{N}E\left(\sum_{i=1}^N u(r_i)\right)+\frac{1}{N^2}E\left[\left(\sum_{i=1}^N u(r_i)\right)^2\right]\\
                                                                & =\frac{1+N}{4N}\\
\end{align*}
And

\begin{align*}
E\left[\frac{1}{N^2}\sum_{i=1}^N u(r_i)\sum_{i=1}^N (1-u(r_i))\right] & =\frac{1}{N}E\left(\sum_{i=1}^N u(r_i)\right)-\frac{1}{N^2}E\left[\left(\sum_{i=1}^N u(r_i)\right)^2\right]\\
                                                                      & =\frac{N-1}{4N}\\
\end{align*}
Then we can get the variance of $C$

\begin{align*}
Var(C) & =\left(\frac{1}{N^+}\right)^2 N^+ \frac{\pi-2+2N^+}{\pi}\sigma^2\frac{1+N}{4N}+\left(\frac{1}{N^-}\right)^2 N^- \frac{\pi-2+2N^-}{\pi}\sigma^2\frac{1+N}{4N}\\
       &+\frac{2}{N^+ N^-}\sqrt{\frac{2}{\pi}}N^+ \sigma\left(-\sqrt{\frac{2}{\pi}}N^- \sigma\right) \frac{N-1}{4N}\\
       & =\left[\frac{(\pi-2)(N+1)}{4\pi N^+ N^-}+\frac{2}{N\pi}\right]\sigma^2\\
\end{align*}

\section{The Distribution of the Statistic $C$}

Denote $X_1=\frac{1}{N^+}\sum_{r^+ \in S^+}r^+$. Because $r^+$ is independent and identically distributed, according to the central limit theorem, $X_1$ is approximately subject to the normal distribution, namely, $X_1 \sim N(\sqrt{\frac{2}{\pi}}\sigma, \frac{\pi-2}{N^+ \pi}\sigma^2)$. Similarly, let $X_2=\frac{1}{N^-}\sum_{r^- \in S^-}r^-$, and $X_2 \sim N(-\sqrt{\frac{2}{\pi}}\sigma, \frac{\pi-2}{N^- \pi}\sigma^2)$.

Let $X_1'=X_1-\sqrt{\frac{2}{\pi}}\sigma$, $X_2'=X_2+\sqrt{\frac{2}{\pi}}\sigma$ and $U'=U-\frac{1}{2}$, the new variables $X_1'$, $X_2'$ and $U'$ are also subject to the normal distribution with the mean $0$. Then the statistic $C$ can be rewritten by

\begin{align*}
C & =X_1U+X_2(1-U)\\
  & =X_2+(X_1-X_2)U\\
  & =X_2+\left(X_1'-X_2'+2\sqrt{\frac{2}{\pi}}\sigma\right)(U'+\frac{1}{2})\\
  & =X_2+2\sqrt{\frac{2}{\pi}}\sigma(U'+\frac{1}{2})+(X_1'-X_2')(U'+\frac{1}{2})\\
\end{align*}

The standard error of $U'$ is $1/\sqrt{4N}$. In our calculation, we chose $N=100$, from which we find that $1/2$ is $10$ times larger than $1/\sqrt{4N}$. The variable $U'$ is subject to the normal distribution with the mean $0$. However, for the normal distribution, the variable falls in the range of $3$ standard error with a probability of $99.7\%$. The constant is $10$ times larger than the standard error, which means that it is far greater than the variable, namely, $1/2>>U'$. Then the equation can be approximated to

\begin{equation*}
C\approx X_2+2\sqrt{\frac{2}{\pi}}\sigma U+\frac{1}{2}(X_1'-X_2')
\end{equation*}
Which is approximately subject to normal distribution.

\section{The Estimation of Standard Error for the Return $r$}

If there is a bubble in time series, the estimated variance $S^2$ would be larger than $\sigma^2$. In order to improve the accuracy of estimation, we remove the data larger than $3$ times standard error, which is equivalent to the estimation from a censored distribution, namely, $E(S'^2\mid |r|\leq 3S)$. If there is no bubble, $S^2$ is the unbalanced estimation of $\sigma^2$, and $E(S'^2\mid |r|\leq 3S)\approx E(S^2\mid |r|\leq 3\sigma)$, then we can get

\begin{align*}
E(S^2\mid |r|\leq 3\sigma) & =\frac{1}{L-1}E\left(\sum_{i=1}^L (r_i-\bar{r})^2\mid |r|\leq 3\sigma\right)\\
                           & =\frac{1}{L-1}E\left(\sum_{i=1}^L r_i^2-L\bar{r}^2\mid |r|\leq 3\sigma\right)\\
                           & =\frac{L}{L-1}\left[E(r^2\mid |r|\leq 3\sigma)-E(\bar{r}^2 \mid |r|\leq 3\sigma)\right]\\
                           & =\frac{L}{L-1}\left[E(r^2\mid |r|\leq 3\sigma)-\frac{1}{L}E(r^2 \mid |r|\leq 3\sigma)\right]\\
                           & =E(r^2\mid |r|\leq 3\sigma)\\
\end{align*}

The conditional distribution of return $r$ can be obtained by

\begin{align*}
F(r\mid |r|\leq 3\sigma) & =P\{R\leq r\mid |r|\leq 3\sigma\}\\
                         & =\frac{P\{-3\sigma\leq R\leq r\}}{P\{-3\sigma\leq R\leq3\sigma\}}\\
                         & =\begin{cases}
                               \frac{\int_{-3\sigma}^r f(r)dr}{2\Phi(3)-1}, & \text{if} \quad |r|\leq 3\sigma\\
                               0, & \text{if} \quad |r|>3\sigma\\
                            \end{cases}
\end{align*}

Where the $\Phi(*)$ means the cumulative distribution function of standard normal distribution. And then we can get the conditional density

\begin{align*}
f(r\mid |r|\leq 3\sigma)=\begin{cases}
                            \frac{f(r)}{2\Phi(3)-1}, & \text{if} \quad |r|\leq 3\sigma\\
                            0, & \text{if} \quad |r|>3\sigma\\
                         \end{cases}
\end{align*}

For the expectation of the variance $S^2$, we can obtain

\begin{align*}
f(S^2\mid |r|\leq 3\sigma) & =E(r^2\mid |r|\leq 3\sigma)\\
                           & =\int_{-\infty}^{+\infty}r^2f(r\mid |r|\leq 3\sigma)dr\\
                           & =\int_{-3\sigma}^{+3\sigma}r^2\frac{f(r)}{2\Phi(3)-1}dr\\
                           & =\frac{1}{2\Phi(3)-1}\int_{-3\sigma}^{+3\sigma}r^2\frac{1}{\sqrt{2\pi} \sigma}e^{-\frac{r^2}{2\sigma^2}}dr\\
                           & =\frac{1}{2\Phi(3)-1}\left(-\frac{6}{\sqrt{2\pi}}e^{-\frac{9}{2}} \sigma^2+\sigma\int_{-3\sigma}^{+3\sigma}\frac{1}{\sqrt{2\pi}}e^{-\frac{r^2} {2\sigma^2}}dr\right)\\
\end{align*}
Let $\mu=r/\sigma$, then it can be rewritten by

\begin{align*}
E(S^2\mid |r|\leq 3\sigma) & =\frac{1}{2\Phi(3)-1}\left(-\frac{6}{\sqrt{2\pi}}e^{-\frac{9}{2}} \sigma^2+\sigma^2\int_{-3}^{+3}\frac{1}{\sqrt{2\pi}}e^{-\frac{1} {2}\mu^2}d\mu\right)\\
                           & =\frac{1}{2\Phi(3)-1}\left(-\frac{6}{\sqrt{2\pi}}e^{-\frac{9}{2}} \sigma^2+(\Phi(3)-\Phi(-3))\sigma^2\right)\\
                           & =\left(1-\frac{6e^{-\frac{9}{2}}}{\sqrt{2\pi}(2\Phi(3)-1)}\right) \sigma^2\\
\end{align*}

\end{document}